\documentclass[prd,aps,nofootinbib,showpacs,showkeys,preprintnumbers,superscriptaddress]{revtex4}
\usepackage{graphicx,amsfonts,amsmath,slashed}
\usepackage{bm}

\begin{document}

\title{Vortex structure of a neutron star with CFL quark core}
\author{D.~M.~Sedrakian}
\email{dsedrak@ysu.am}
\affiliation{Department of Physics, Yerevan State University, Alex
        Manoogian Str. 1, 375025 Yerevan, Armenia}
\author{D.~Blaschke}
\email{blaschke@ift.uni.wroc.pl}
\affiliation{
Institute for Theoretical Physics, University of Wroclaw,\\
Max Born pl. 9,
50-204 Wroclaw, Poland}
\affiliation{Bogoliubov Laboratory for Theoretical Physics,
Joint Institute for Nuclear Research,\\ 
141980 Dubna, Moscow Region, Russia}
\author{K.~M.~Shahabasyan}
\email{kshahabas@ysu.am}
\affiliation{Department of Physics, Yerevan State University, Alex
        Manoogian Str. 1, 375025 Yerevan, Armenia}
\author{M.~K.~Shahabasyan}
\email{mshahabas@ysu.am}
\affiliation{Department of Physics, Yerevan State University, Alex
        Manoogian Str. 1, 375025 Yerevan, Armenia}

\begin{abstract}
The Ginzburg-Landau equations for magnetic and gluomagnetic gauge fields of 
the semi-superfluid strings in the color superconducting core of the neutron 
star with diquark condensate in the color-flavor-locking (CFL) phase 
are derived. 
The simultaneous coupling of the CFL diquark condensate to the magnetic and 
gluo­magnetic gauge fields is taken into account. 
The asymptotic values of these string energies are defined from quantization 
conditions. 
It is shown that a lattice of semi-superfluid strings with minimal quantum 
of circulation emerges in the superconducting quark core during rotation of 
the star. The magnetic field in the core of this string is of order of 
$10^{18}$ G. 
The cluster of proton vortices appeared in the hadronic phase  around each 
superfluid neutron vortex due to entrainment effect, creates in the 
superconducting quark core new semi-superfluid strings with minimal quantum of 
circulation.
\end{abstract}
\maketitle


\section{Introduction}

At present the existence of different types of quantum vortex lattices in 
different phases of neutron star matter is confirmed to a high degree of
confidence.
The presence of superfluid neutron vortex lattices in the inner crust 
(``A~e~n'' - phase) and in the hadronic core (``n~p~e'' - phase) as a 
consequence of the star rotation \cite{Ginzburg:1964} explains the 
peculiarities of the rotational dynamics of pulsars: glitches of the pulsar
angular velocity, post-glitch relaxation 
\cite{Baym:1969,Anderson:1975,Alpar:1984,Jones:1990} 
and quasi-sinusoidal oscillations of the angular velocity
\cite{Ruderman:1970,Sedrakian:1995}.

The protons in the hadronic phase represent a type-II superconductor and 
therefore they accumulate the relict magnetic field of the neutron star in 
magnetic quantum vortex lattices with the flux $\Phi_0=\pi\hbar c/e$ 
\cite{Baym:1969a}.
The presence of the strong interaction between the neutron and proton 
condensates leads to an entrainment of the superconducting protons by the 
rotating superfluid neutrons \cite{Sedrakian:1980,Vardanyan:1981}. 
Due to the entrainment effect arises a nonvanishing proton current around a 
neutron vortex which leads to the appearance of a magnetic flux $\Phi_1$ 
at the neutron vortex which is not a multiple of the flux quantum $\Phi_0$ 
\cite{Sedrakian:1980a,Alpar:1984a} 
and a magnetic entrainment field strength which generates new proton vortices 
with the flux $\Phi_0$ \cite{Sedrakian:1982}.
In the region around the neutron vortex where $H(r)>H_{c1}$ appears an 
inhomogeneous cluster of proton vortices. The mean induction of the magnetic
field of the neutron star that originates from these clusters is of the
order of $10^{12}$ G. \cite{Sedrakian:1983,Sedrakian:1991}. 
The influence of these clusters on the rotational dynamics of pulsars has 
been considered in \cite{Sedrakian:1995a}. 
Note that the entrainment effect in the $^3He$ - $^4He$ mixture has been 
studied in the work  \cite{Andreev:1975}.

In the superdense core of the star the presence of quark matter is possible.
In this matter due to the attractive interaction between quarks in the color
antitriplet channel, caused by one-gluon exchange, one expects the appearance
of a superconducting diquark condensate.
The quark pairs which form the condensate have zero total angular momentum
$J$ \cite{Bailin:1984,Alford:1998,Rapp:1998}.
It has been shown that in chiral quark models with nonperturbative four-point
interactions originating from instantons \cite{Carter:1999}
or nonperturbative gluon propagators \cite{Blaschke:1998} the anomalous quark 
pairing amplitudes are of the order of $100$ MeV.
Consequently one may suggest that the diquark condensate will exist at 
densities exceeding the deconfinement density and at temperatures below the
critical $T_c$ of the order of $50$ MeV, see, e.g., 
Ref.~\cite{Blaschke:2005uj} for a phase diagram.

There are two types of scalar condensates with large gaps of the order of
$100$ MeV: the {\it 2SC phase} \cite{Bailin:1984}, where only $u$ and $d$
quarks of two colors pair and the {\it CFL phase}, where approximately 
massless $u$, $d$ and $s$ quarks of all three colors pair
\cite{Alford:1999,Schafer:1999,Schafer:2000}. 
The {\it CFL phase} with three flavors of massless quarks is the most 
stable one in the weak coupling limit at $T=0$ and in the vicinity of the
critical temperature $T_c$.
In the work \cite{Blaschke.1999} it has been shown that the diquark condensate 
in the {\it 2SC phase} is a type-II superconductor.
The presence of electric and color charges of the Cooper pair leads to the
appearance of electric and color superconductivity in the {\it 2SC} and
{\it CFL} phases.
These two phenomena are not independent because photon and gluon gauge fields 
are coupled to each other. 
One of the resulting mixed fields is massless while the other obtains a mass
\cite{Alford:2000}.
It has been conjectured \cite{Berdermann:2006rk} that the occurrence of a 
{\it CFL} phase in newly born magnetized strange stars can give rise to 
asymmetric neutrino propagation and thus bear interesting consequences for 
gamma-ray bursts and pulsar kicks. 

The Ginzburg-Landau equations for the {\it 2SC} phase accounting for the 
rotated fields have been obtained in \cite{Blaschke:2000}.
These equations were used in \cite{Sedrakian:2001} for the study of the action
of an external homogeneous magnetic field on the superconducting quark core of 
a neutron star. 
It was shown that in the absence of vortex lattices Meissner currents in the 
core screen the magnetic field almost completely. 
In Ref. \cite{Sedrakian:2002} a magnetic field configuration with 
superconducting {\it 2SC} quark core has been found in which the magnetic 
field is generated in the hadronic {\it ``n~p~e''}- phase due to the 
effect of entrainment of superconducting protons by the superfluid neutrons  
\cite{Sedrakian:1991} and penetrates into the quark core in the form of 
quark magnetic vortex lattices which form due to the presence of screening
Meissner currents.

The Ginzburg-Landau (GL) free energy of the homogeneous superconducting 
{\it CFL} phase has been found in \cite{Schafer:2000,Iida:2001}.
A gauge invariant expression for the GL free energy has been obtained in
\cite{Gorbar:2000}. 
The GL equations in the presence of an external field have been formulated in 
\cite{Iida:2002,Giannakis:2003}.
We remark that the {\it CFL} condensate has superconducting as well as 
superfluid properties.
This is due to the breaking of local symmetries - color $SU(3)_c$ and 
electromagnetic $U(1)_{\rm EM}$, as well as global symmetries -
flavor $SU(3)_F$ and baryon number $U(1)_B$.
Therefore naturally singular solutions appear: magnetic vortex lattices and 
superfluid vortex lattices. In the works \cite{Iida:2002,Forbes:2002},
superfluid quark vortex lattices have been considered, based on the breaking 
of the global $U(1)_B$ symmetry.
In Ref.~\cite{Iida:2005}, using the point of view provided in 
\cite{Giannakis:2003}, abelian magnetic vortex lattices have been studied
which are based on the current of a massive gauge field.

Finally in Ref.~\cite{Balachandran:2006}, on the basis of a topological and 
group-theoretical analysis of the GL free energy of the {\it CFL} phase,
new nonabelian semi-superfluid vortices $M_1$ and $M_2$ have been
found which have properties of superfluid vortices as well as magnetic 
vortices. 
In contrast to superfluid and magnetic vortices, these vortices are 
topologically stable. 
One superfluid quark $U(1)_B$ vortex is topologically equivalent to three 
nonabelian semi-superfluid vortices $M_1$.
In \cite{Nakano:2007} the interaction between two parallel semi-superfluid 
vortices has been studied.
It has been shown that between two vortices act long-range repulsive forces,
and attractive forces between two antivortices.
Conclusions have been drawn about the possibiity of the decay of of the 
superfluid   $U(1)_B$ vortex to three semi-superfluid $M_1$ vortices and about
the possibility of the existence of a stable $M_1$ vortex lattice. 

The aim of the present work is the study of the vortex structure and magnetic 
field of a neutron star accounting for the presence of a {\it CFL} quark 
core with semi-superfluid vortices.
Moreover, we account for the rotation of the star and the occurrence of a 
magnetic induction in the hadronic phase due to the entrainment effect.
We do not consider the fossil magnetic field. 

\section{Vortex lattices in the CFL phase}

In the homogeneous CFL phase, in the absence of fields the order parameter
$\Phi$ is determined as
\begin{equation}
        \label{Phi}
        \Phi = |k_A|\left(
        \setlength\arraycolsep{-0.01cm}
        \begin {array}{ccc}
        1&~0&~0\\
        0&~1&~0\\
        0&~0&~1\\
       \end {array}
        \right)~,
\end{equation}
where $k_A=(-\alpha/8\beta)^{1/2}$.
The order parameter $\Phi_B$ of the superfluid $U(1)_B$ vortex has the form
\begin{equation}
        \label{PhiB}
        \Phi_B = |k_A| e^{i\vartheta}
        \left(
        \setlength\arraycolsep{-0.01cm}
        \begin {array}{ccc}
        1&~0&~0\\
        0&~1&~0\\
        0&~0&~1\\
       \end {array}
        \right)~,
\end{equation}
where $\vartheta$ is the polar angle of the radius vector in the $x-y$ plane.
The nonabelian semi-superfluid vortex $M_1$ which has been considered in
\cite{Balachandran:2006} is described by the following order parameter
\begin{equation}
        \label{Phi1}
        \Phi_1 = |k_A| 
        \left(
        \setlength\arraycolsep{-0.01cm}
        \begin {array}{ccc}
        f(r)e^{i\vartheta}&~0&~0\\
        0&~1&~0\\
        0&~0&~1\\
       \end {array}
        \right)~,
\end{equation}
The order parameter of the second nonabelian semi-superfluid vortex $M_2$ is
\begin{equation}
        \label{Phi2}
        \Phi_2 = |k_A| 
        \left(
        \setlength\arraycolsep{-0.01cm}
        \begin {array}{ccc}
        f(r)e^{i 2\vartheta}&~0&~0\\
        0&~1&~0\\
        0&~0&~1\\
       \end {array}
        \right)~.
\end{equation}
Note that our definition of the order parameter $ \Phi_2 $ differs from that
of Ref.~\cite{Balachandran:2006}.  
The condition for quantization of the superfluid $U(1)_B$ vortex is given by
\cite{Iida:2002}
\begin{equation}
\frac{2}{3}m_B\oint d\vec{l} \vec{v}_q=2 \pi \hbar~,
\end{equation}
where $m_B$ is the baryon mass. From this equation we get the expression for
the velocity $v_q$
\begin{equation}
v_q=\frac{3\hbar}{2m_B}\frac{1}{r}~,
\end{equation}
The total kinetic energy energy $E_q$ of the superfluid  $U(1)_B$ vortex per
length unit (linear tension)
\begin{eqnarray}
E_q&=&\frac{1}{2}\int_0^{2\pi} d\vartheta \int_\xi^R dr~m_B n_S v_q^2
\nonumber\\
&=&\varrho \frac{\kappa_B^2}{4\pi} \ln\frac{R_q}{\xi} ~,
\end{eqnarray}
where $\varrho=m_B~n_S$ is the mass density of the superfluid
component with the number density $n_S=16 K_Tm_B |k_A|^2/3$. 
$R_q$ is the radius of the $U(1)_B$ vortex and $\xi$ the correlation length of 
the diquark pair or the characteristic redius of the normal core of the vortex.
The quantum of circulation is $\kappa_B=3\pi\hbar/m_B$, see below.

Correspondingly, the expression for the asymptotic velocity of the nonabelian
semi-superfluid vortex $M_1$ is obtained from the following quantization
condition
\begin{eqnarray}
\label{quant-v}
\frac{2}{3}m_B\oint d\vec{l} \vec{v}_q= \frac{2}{3}\pi n \hbar~,~~~
v_{\rm 1S}=\frac{\hbar}{2m_B}\frac{1}{r}~,~(n=1)~,
\end{eqnarray}
and for the vortex  $M_2$ we obtain (for $n=2$)
\begin{equation}
v_{\rm 2S}=\frac{\hbar}{m_B}\frac{1}{r}~.
\end{equation}
For the asymptotic expression of the linear tension of the semi-superfluid 
vortex $M_1$ we obtain
\begin{eqnarray}
\label{E1S}
E_{\rm 1S}&=&
\frac{1}{2}\int_0^{2\pi} d\vartheta\int_\xi^R dr~r~m_B n_S v_{\rm 1S}^2
\nonumber
\\
&=&\varrho \frac{\kappa_1^2}{4\pi} \ln\frac{R_1}{\xi} ~,
\label{E2S}
\end{eqnarray}
where $\kappa_1=\pi\hbar/m_B$ is the quantum of circulation and $R_1$ the 
radius of the semi-superfluid vortex $M_1$.
This expression coincides with the result obtained in \cite{Balachandran:2006}.
The asymptotic expression of the linear tension of the semi-superfluid 
vortex $M_2$ is equal to
\begin{eqnarray}
E_{\rm 2S}&=&
\frac{1}{2}\int_0^{2\pi} d\vartheta\int_\xi^R dr~r~m_B n_S v_{\rm 2S}^2
\nonumber\\
&=&\varrho \frac{\kappa_1^2}{\pi} \ln\frac{R_2}{\xi} ~,
\end{eqnarray}
It also coincides with the asymptotic energy of the vortex $M_2$ in Ref.
\cite{Balachandran:2006}.
We point out that these asymptotic energies have been obtained in 
\cite{Balachandran:2006} from a gradient-invariant expression for the GL 
kinetic energy in which the massive gauge field  $\vec{A}_x$ has been 
considered.
From the Equations (5) and (8) follows that the quanta of the circulation 
of the superfluid $U(1)_B$ vortex and of the semi-superfluid $M_1$ vortex
are $\kappa_B  $ and $\kappa_1$, respectively.
Since the quantum of circulation  $\kappa_B  $ is three times larger than 
the quantum of circulation $\kappa_1 $ and between the semi-superfluid 
vortices act repulsive long-range forces \cite{Nakano:2007}, 
each superfluid  $U(1)_B$ vortex which occurs in the CFL phase decays 
into three vortices  $M_1$.
Consequently, in the CFL quark core of a neutron star under certain 
external conditions can occur a stable lattice of semi-superfluid vortices 
 $M_1$. From Eq. (9) follows that the quantum of circulation of the 
semi-superfluid vortex   $M_2$ is equal to  $\kappa_2=2\pi \hbar / m_B$.
If we write the quatization condition (\ref{quant-v}) in the canonical form, 
e.g., as circulation of momentum equal to $2\pi \hbar n$, then the new 
quantum number for the semi-superfluid vortex $M_1$ is $n=1/3$ and for the 
semi-superfluid vortex $M_2$ it is $n=2/3$.


The superfluid $U(1)_B$ vortices and the semi-superfluid vortices 
$M_1$ and $M_2$ can occur in the quark core due to the rotation of the star.
The critical angular velocity for the appearance of the $U(1)_B$ vortex is
\cite{Iida:2002}
\begin{equation}
\label{omega-c1}
\omega_{c1}^B=\frac{3\hbar }{2m_B R_q^2}\ln \frac{R_q}{\xi}~,
\end{equation}
where $R_q$ is the radius of the quark core.
For the determination of the critical angular velocity $\omega_{c1}^\prime$
for the occurrence of the semi-superfluid vortex $M_1$ we find its angular
momentum $L_{1S}$
\begin{equation}
\label{L1S}
L_{\rm 1S}=\int_0^{2\pi}d\vartheta \int_\xi^{R_q}dr r m_B n_S 
{\mathbf r}{\mathbf v}_{\rm 1S}
=\frac{1}{2}\varrho \kappa_1 R_q^2~.
\end{equation}
We have omitted a second term on the right hand side of eq. (\ref{L1S}) because
it is proportional to $\xi^2$, where $\xi\approx 0.76$ fm 
\cite{Sedrakian:2007}. 
Further, from the condition $E_{\rm 1S}-\omega_{c1}^\prime L_{\rm 1S}=0$
we find 
\begin{equation}
\label{om-c1p}
\omega_{c1}^\prime=\frac{\hbar }{2m_B R_q^2}\ln \frac{R_q}{\xi}~.
\end{equation}
The angular momentum $L_{2S}$ of the semi-superfluid vortex $M_2$ is determined
as
\begin{equation}
\label{L2S}
L_{\rm 2S}=\int_0^{2\pi}d\vartheta \int_\xi^{R_q}dr r m_B n_S
{\mathbf r}{\mathbf v}_{\rm 2S}
=\frac{1}{2}\varrho \kappa_2 R_q^2~.
\end{equation}
The critical angular velocity $\omega_{c1}^{\prime \prime}$
for the occurrence of the semi-superfluid vortex $M_2$ we find from 
$E_{\rm 2S}-\omega_{c1}^{\prime\prime} L_{\rm 2S}=0$ as
\begin{equation}
\label{om-c1pp}
\omega_{c1}^{\prime\prime}=\frac{\hbar }{m_B R_q^2}\ln \frac{R_q}{\xi}~.
\end{equation}
Correspondingly, upon rotation of the star in the CFL phase for an angluar 
velocity $\omega > \omega_{c1}^\prime$ appears a lattice of semi-superfluid 
vortices $M_1$.
This lattice will be stable because between the semi-superfluid vortices $M_1$
act long-range repulsive forces \cite{Nakano:2007}.
For the determination of their density $n_v$ we find the circulation of the 
superfluid velocity $\vec{v}_{\rm 1S}$ along the contour $L$ that goes along 
the border of the quark core
\begin{eqnarray}
\label{circ-v}
\oint_{L} d\vec{l} \vec{v}_{1S}= 
\int_{S} d\vec{S} {\rm rot}\vec{v}_{1S}= 2\omega \pi R_q^2=\kappa_1N_q~,
\end{eqnarray}
where $N_q$ is the total number of $M_1$ vortices. 
From this follows for the density of vortices $n_v=2\omega/\kappa_1$.

The superfluid $U(1)_B$ vortices appearing in the CFL phase upon star 
rotation are unstable, because they are topologically equivalent to three 
$M_1$ vortices. Their energy $E_q$ and critical angular velocity 
$\omega_{c1}^B$ are $9$ times and $3$ times larger than the  
 energy $E_{\rm 1S}$ and critical angular velocity $\omega_{c1}^\prime$, 
respectively, and between the $M_1$ vortices act long-range repulsive forces.
Consequently, the superfluid $U(1)_B$ vortices decay to three 
semi-superfluid vortices $M_1$. 
Correspondingly, the semi-superfluid vortices $M_2$ appearing in the CFL phase
upon rotation decay to two $M_1$ vortices.

For a rotating star in the hadronic phase for the values 
$\omega>\omega_{c1}^n=\hbar/(2m_B~R_n^2)\ln(R_n/\xi_n)$ 
arises a lattice of superfluid neutron vortices with the quantum of circulation
$\kappa_n=\pi \hbar/m_B$ and with the density $n_n=2\omega/\kappa_n$.
Assuming a radius of the hadronic core $R_n=5\cdot 10^5$ cm and a coherence
length for neutrons $\xi_n=3.1\cdot 10^{-12}$ cm we get 
$\omega_{c1}^n=5\cdot 10^{-14}$ s$^{-1}$.
Calculating the critical angular velocity of the appearance of semi-superfluid 
vortices $M_1$ according to the formula (\ref{om-c1p})  
we get for a radius of the quark core  $R_q= 10^5$ cm the result
 $\omega_{c1}^\prime=1.3\cdot 10^{-12}$ s$^{-1}$.
Since the angular velocities of the rotation of pulsars are by far exceeding
the critical angular velocity $\omega_{c1}^\prime$, in pulsars with a quark 
core there occurs a lattice of semi-superfluid vortices $M_1$ with a density 
$n_v=10^3~\omega$ cm$^{-2}$.
Because the quanta of circulation and the densities of both vortices are equal,
these semi-superfluid vortices merge at the border of the hadronic and the 
CFL quark matter phases with the superfluid neutron vortices.
The baryonic chemical potential is continuous at the border also.
Note that superfluid neutron vortices and semi-superfluid quark vortices $M_1$ 
form a triangular lattice with the constant 
$d=(\pi\hbar/(\sqrt{3} m_B) \omega)^{1/2}$.
The value $d$ for pulsars is of the order $10^{-3}$ cm.

\section{Ginzburg-Landau equations (GLE) for magnetic and gluomagnetic fields 
of semi-superfluid vortices}

In the work \cite{Balachandran:2006} on the basis of a topological analysis
has been formulated the following quantization condition for the massive
gauge field $\vec{A}_x$
\begin{equation}
\label{q_x}
q_x \oint_L d\vec{l} \vec{A}_x = 2 \pi \hbar c~,
\end{equation}
where $q_x=\sqrt{3g^2+4e^2}/2$. We remark that $q_x=3 q_{\rm CFL}$, where 
$q_{\rm CFL}$ is the quantity named mixed charge, introduced in 
\cite{Iida:2005}. The mixed fields $\vec{A}_x$ and  $\vec{A}_y$ are defined 
in the following way \cite{Gorbar:2000}
\begin{equation}
\label{mixedfield}
\vec{A}_x =\vec{A}\sin \alpha + \vec{A}^8 \cos \alpha ~,~~~
\vec{A}_y =\vec{A}\cos \alpha - \vec{A}^8 \sin \alpha ~,
\end{equation}
where $\cos \alpha = \sqrt{3} g/ \sqrt{3g^2+4e^2}$, and $\vec{A}$ is the 
vector potential of the magnetic field while $\vec{A}^8$ is the 
vector potential of gluomagnetic field. 
Here, $g$ is the strong interaction coupling constant 
$(g^2/ (4 \pi \hbar c)\approx 1)$ and $e$ is the constant of the
electromagnetic interaction $e^2/ (4 \pi \hbar c \approx 1/137)$.

Then quantization condition (\ref{q_x}) is written in the form
\begin{equation}
e \int d\vec{S}~{\rm rot} \vec{A}+\frac{\sqrt3}{2} g\int d\vec{S}~{\rm rot} 
\vec{A}^8=2\pi \hbar c~.
\end{equation}
Further, when we insert in Eqs. (12) and (13) of the work \cite{Sedrakian:2007}
the expressions $\Phi(r)=|k_A|f(r)e^{i\vartheta}$, $\chi(r)=\sqrt{2}|k_A|$, 
$q=q_x/3$, we get the GLE which describe the magnetic and gluomagnetic fields
of the semi-superfluid vortex $M_1$ 
\begin{eqnarray}
\label{rotrotA}
\lambda_q^2 {\rm rot~rot}\vec{A} + \vec{A} \sin^2 \alpha 
&=& \frac{3 f^2(r)\sin \alpha \nabla \vartheta}{q_x [2 f^2(r)+1]}
-\vec{A}^8 \sin\alpha \cos\alpha~,
\\
\lambda_q^2 {\rm rot~rot}\vec{A}^8+ \vec{A}^8 \cos^2\alpha 
&=& \frac{3 f^2(r) \cos \alpha ~\nabla \vartheta}{q_x [2 f^2(r)+1]} 
- \vec{A} \sin \alpha \cos \alpha~,
\label{rotrotA8}
\end{eqnarray}
where the penetration depth $\lambda_q$ is given by
\begin{equation}
\lambda_q^{-2}=\frac{32\pi}{9}K_Tq_x^2 |k_A|^2 [2 f^2(r)+1]~.
\end{equation}
Since at large distances from the vortex $f(r)\to 1$, the Eqs. 
(\ref{rotrotA}) and (\ref{rotrotA8}) can be
rewritten in the form
\begin{eqnarray}
\label{rotrotA-2}
\lambda_q^2 {\rm rot~rot}\vec{A} + \vec{A} \sin^2 \alpha 
&=& \frac{\Phi_x\sin \alpha \nabla \vartheta}{2\pi}
-\vec{A}^8 \sin\alpha \cos\alpha~,
\\
\lambda_q^2 {\rm rot~rot}\vec{A}^8+ \vec{A}^8 \cos^2\alpha 
&=& \frac{\Phi_x \cos \alpha ~\nabla \vartheta}{2\pi} 
- \vec{A} \sin \alpha \cos \alpha~.
\label{rotrotA8-2}
\end{eqnarray}
Here the London penetration depth $\lambda_q$ and the flux of the massive field
$\Phi_x$ are given by $\lambda_q^{-2}={32\pi}K_Tq_x^2 |k_A|^2/3$ and 
$\Phi_x=2\pi\hbar c/q_x$.
Note that the London penetration depth is $\lambda_q\approx 1$ fm 
\cite{Sedrakian:2007}.
We integrate Eq. (\ref{rotrotA-2}) along the contour $L$ which goes along the 
border of the quark core region. Since the current at the border vanishes therefore
\begin{equation}
\label{int-rotrot}
\oint_L d\vec{l} {\rm rot~rot} \vec{A} =\oint_L d\vec{l} {\rm rot} \vec{B}
=0~.
\end{equation}
The condition of confinement is written as
\begin{equation}
\label{conf}
\oint_L d\vec{l} \vec{A}^8 = 0~.
\end{equation}
The total flux of the magnetic field through the core is equal to
\begin{equation}
\label{Phi-tot}
\oint_L d\vec{l} \vec{A} = 
\oint_S d\vec{S} \vec{B} = \frac{\Phi_x}{\sin \alpha}N_q~,
\end{equation}
where $N_q$ is the total number of semi-superfluid vortices $M_1$ in the core.
Using the relation $\sin \alpha = 2e/ \sqrt{3g^2+4e^2}$ we get the following 
expression for the flux of the magnetic field $\Phi_M$ of the 
semi-superfluid vortex $M_1$
\begin{equation}
\label{PhiM}
\Phi_M = 2\pi\hbar c/e=2\Phi_0~,
\end{equation}
where $\Phi_0=2\cdot 10^{-7}$ Gcm$^2$ is the magnetic flux quantum.
This way these vortices have a doubled flux quantum.
Note that the expression for the flux is a consequence of the condition of the 
uniqueness of the wave function
\begin{equation}
\label{uni}
\oint_L d\vec{l} \nabla \vartheta = 2\pi~,
\end{equation}
which follows from the quantization condition (\ref{quant-v}). 

Further, when we insert in Eqs. (12) and (13) of the work \cite{Sedrakian:2007}
the expressions $\Phi(r)=|k_A|f(r)e^{i2\vartheta}$, $\chi(r)=\sqrt{2}|k_A|$, 
$q=q_x/3$, we get the GLE which describe the magnetic and gluomagnetic fields
of the semi-superfluid vortex $M_2$ 
\begin{eqnarray}
\label{rotrotA-3}
\lambda_q^2 {\rm rot~rot}\vec{A} + \vec{A} \sin^2 \alpha 
&=& \frac{6 f^2(r)\sin \alpha \nabla \vartheta}{q_x [2 f^2(r)+1]}
-\vec{A}^8 \sin\alpha \cos\alpha~,
\\
\lambda_q^2 {\rm rot~rot}\vec{A}^8+ \vec{A}^8 \cos^2\alpha 
&=& \frac{6 f^2(r) \cos \alpha ~\nabla \vartheta}{q_x [2 f^2(r)+1]} 
- \vec{A} \sin \alpha \cos \alpha~,
\label{rotrotA8-3}
\end{eqnarray}
where the penetration depth $\lambda_q$ is given by
\begin{equation}
\lambda_q^{-2}=\frac{32\pi}{9}K_Tq_x^2 |k_A|^2 [2 f^2(r)+1]~.
\end{equation}
At large distances from the vortex, the Eqs. 
(\ref{rotrotA-3}) and (\ref{rotrotA8-3}) can be
rewritten as
\begin{eqnarray}
\label{rotrotA-4}
\lambda_q^2 {\rm rot~rot}\vec{A} + \vec{A} \sin^2 \alpha 
&=& \frac{2 \Phi_x\sin \alpha \nabla \vartheta}{2\pi}
-\vec{A}^8 \sin\alpha \cos\alpha~,
\\
\lambda_q^2 {\rm rot~rot}\vec{A}^8+ \vec{A}^8 \cos^2\alpha 
&=& \frac{2 \Phi_x \cos \alpha ~\nabla \vartheta}{2\pi} 
- \vec{A} \sin \alpha \cos \alpha~.
\label{rotrotA8-4}
\end{eqnarray}
Correspondingly, at the border of the quark core the $M_2$ vortices have 
the magnetic flux  $4\Phi_0$.

We will now consider the GLE for the rotated field of the vortices introduced above.
For the rotated field of the semi-superfluid vortex $M_1$ 
we have \cite{Balachandran:2006}
\begin{equation}
\label{dBxdr}
-\hat{e}_\vartheta \frac{d B_x^z}{dr}
=\frac{32\pi}{3c} K_Tq_x |k_A|^2
\left\{ \hbar f^2(r)~\nabla \vartheta 
-\frac{q_x}{3c} [2 f^2(r)+1] \vec{A}_\vartheta\right\}~,
\end{equation}
where $\vec{A}_\vartheta=\hbar c \gamma(r)/(q_xr)\hat{e}_\vartheta$
is the only nonvanishing component of the potential $\vec{A}_x$,
$B_x^z$ is the $z$- component of the massive vector field 
$B_x={\rm rot} \vec{A}_x$ and $\hat{e}_\vartheta$ is the unit vector 
of polar coordinates.
The Maxwell equation for the massive vector field (\ref{dBxdr}) 
can be represented as
\begin{equation}
\label{Bx}
-\frac{d B_x^z}{dr}= \frac{4\pi}{c}j_{\rm 1x}~,
\end{equation}
where $j_{\rm 1x}=n_{\rm 1S}q_x {\bf v}_{\rm 1x}$ is the current density.
Here the number density $n_{\rm 1S}$ and the ``superfluid'' velocity
${\bf v}_{\rm 1x}$ are determined by
\begin{eqnarray}
n_{\rm 1S}&=&\frac{16\pi}{3} K_T m_B|k_A|^2 f^2(r)~,
\\
{\bf v}_{\rm 1x}&=&\frac{\hbar}{2m_B}\frac{1}{r}
- \frac{q_x}{m_Bc}\left(\frac{1}{3}+\frac{1}{6 f^2(r)}\right)
\vec{A}_\vartheta~.
\end{eqnarray}
It follows that the quantum of circulation of this vortex is again equal to
the quantum of circulation of the superfluid neutron vortex.
Since at large distances $\gamma(r)\to 1$ and $f(r)\to 1$, the 
``superfluid'' velocity ${\bf v}_{\rm 1x}$ and the current $j_{\rm 1x}$
go to zero, and the density is $n_{\rm 1S}=n_{\rm S}$.

For the abelian ``magnetic'' vortex $A_x$, considered in 
Ref.~\cite{Nakano:2007}, the Maxwell equation is  
\begin{equation}
\label{dBxdr-abel}
-\hat{e}_\vartheta \frac{d B_x^z}{dr}
=\frac{4\pi}{c} \left\{\frac{16}{3} K_T m_B q_x |k_A|^2
\left(\frac{3\hbar}{2m_B}\nabla \vartheta 
-\frac{q_x}{2m_B c}\vec{A}_\vartheta\right)\right\}~.
\end{equation}
From Eq.~(\ref{dBxdr-abel}) follows that the quantum of circulation of the 
abelian magnetic vortex is equal to $\kappa_B$, i.e. three times larger than 
the quantum of circulation $\kappa_1$.
We note also that for $r\to \infty$ the current $\vec{j}_x$ goes to zero for 
the vector potential 
$\vec{A}_\vartheta=3\hbar c \gamma(r)/(q_xr)\hat{e}_\vartheta$, corresponding
to a flux of the massive field $3\Phi_x$.
Therefore these vortices are unstable and will decay into three $M_1$ vortices.

For the vortex $M_2$ the Maxwell equation has the following form
\begin{equation}
\label{dBxdr-massive}
-\hat{e}_\vartheta \frac{d B_x^z}{dr}
=\frac{4\pi}{c} \left\{\frac{16}{3} K_T m_B q_x |k_A|^2 f^2(r)
\left[\frac{\hbar}{m_B}\nabla \vartheta 
-\frac{q_x}{m_B c}\left(\frac{1}{3}+\frac{1}{6 f^2(r)} \right)
\vec{A}_\vartheta\right]\right\}~.
\end{equation}
From Eq.~(\ref{dBxdr-massive}) follows the expression for the velocity
\begin{equation}
\label{vx-massive}
{\bf v}_{\rm 2x}=\frac{\hbar}{m_B}\frac{1}{r}
- \frac{q_x}{m_Bc}\left(\frac{1}{3}+\frac{1}{6 f^2(r)}\right)
\vec{A}_\vartheta~.
\end{equation}
The corresponding quantum of circulation is $\kappa_2=2\pi\hbar / m_B$.
We remark that for $r\to \infty$ the velocity ${\bf v}_{\rm 2x}$ and 
the current $\vec{j}_x$ go to zero for the vector potential 
$\vec{A}_\vartheta=2\hbar c \gamma(r)/(q_xr)\hat{e}_\vartheta$, which
corresponds to the flux of the massive field $2\Phi_x$.

So the Maxwell equations for nonabelian semi-superfluid vortices $M_1$ and 
$M_2$ confirm our classification of vortices by their quanta of 
circulation and our conclusion that $M_1$ vortices will be only stable 
ones in the quark core.

The expression for the massive field in the center of the vortex $M_1$ 
from the GLE for the order parameter is
\begin{equation}
\label{Bx0}
B_x(0)=\frac{3\Phi_x}{4\pi \xi^2}=1\cdot 10^{18}~{\rm G}~.
\end{equation}
We notice that the above critical field for the destruction of the 
superconductivity is $H_{c2}=\Phi_x/(2\pi\xi^2)=1.05\cdot10^{18}~G.$

As it was mentioned earlier, due to the effect of entrainment inside each
neutron vortex in the hadronic phase occurs a cluster of proton vortices with 
the radius $\delta_n=10^{-5}$ cm, which contains about $10^{12}$ vortices.
The mean magnetic induction of a cluster is of the order of $10^{14}$ G
\cite{Sedrakian:1983,Sedrakian:1995a}.
Due to conservation of the magnetic field the proton vortex cluster 
creates around the original semi-superfluid vortex $M_1$ with the radius 
$R=10^{-3}$ cm new semi-superfluid vortices $M_1$ with the radius 
$\lambda_p=10^{-11}$ cm, whereby two proton vortices with flux $\Phi_0$ each
merge at the border of the hadronic and CFL phases into one new semi-superfluid
vortex   $M_1$.

\section{Conclusion}

We have obtained the GLE for the magnetic and gluomagnetic gauge 
fields of the nonabelian semi-superfluid vortex lattices in the color 
superconducting core of a neutron star which contains a {\it CFL} diquark
condensate.
We have determined the asymptotic values of the energy of these vortex 
lattices and the critical angular velocities of their appearance from the 
quantization condition.
The quantum  numbers in this condition for nonabelian semi-superfluid vortices
are non-integer.
We have classified all possible vortex lattices in the  {\it CFL} phase
according to their quanta of circulation.
We have shown that for a rotating star in the superconducting quark core 
appears a lattice of semi-superfluid vortices $M_1$ which has the smallest 
quantum of circulation. In the hadronic phase appears a stable lattice of 
superfluid neutron vortices,
each of which joining at the phase border with one of the semi-superfluid 
vortices  $M_1$.
The clusters of proton vortices that appear in the hadronic phase around each 
superfluid neutron vortex due to the entrainment effect, generate in the quark
core new semi-superfluid vortex lattices $M_1$.

\section*{Acknowledgements}
M.K.S. acknowledges support from the Helmholtz International Summer Schools 
(HISS) programme which gave the possibility to start this work 
during the meeting on {\it Dense Matter in Heavy-Ion Collisions and 
Astrophysics} in  Dubna, Russia, July 2008. 
D.B. is grateful to the Volkswagen Foundation and to the European Science 
Foundation Research Networking Programme {\it CompStar} for supporting his 
participation at the International Workshop on {\it The Modern Physics of 
Compact Stars} in Yerevan, Armenia, September 2008, where this work has been 
completed.

\end{document}